\documentclass[%
 showpacs,
 showkeys,
 amsmath,amssymb,
]{revtex4-1}

\usepackage{graphicx}
\usepackage{dcolumn}
\usepackage{bm}

\usepackage{CJK}

\usepackage{color}
\usepackage{amssymb}
\usepackage{amsfonts}

\usepackage[colorlinks,urlcolor=blue,linkcolor=blue,citecolor=blue,anchorcolor=blue]{hyperref}

\begin{document}


\preprint{APS/123-QED}

\title{Asymmetric conical diffraction in dislocated edge-centered square lattices}

\author{Hua Zhong$^{1,2,3}$}
\author{Rong Wang$^{1,2,3}$}
\author{Milivoj R. Beli\'c$^{4}$}
\author{Yanpeng Zhang$^{1,2}$}
\author{Yiqi Zhang$^{1,2,3}$}
\email{zhangyiqi@mail.xjtu.edu.cn}

\affiliation{%
 $^1$Key Laboratory for Physical Electronics and Devices of the Ministry of Education \& Shaanxi Key Lab of Information Photonic Technique,
Xi'an Jiaotong University, Xi'an 710049, China \\
$^2$School of Electronic and Information Engineering, Xi'an Jiaotong University, Xi'an, Shaanxi 710049, China\\
$^3$Guangdong Xi'an Jiaotong University Academy, Foshan 528300, China\\
$^4$Science Program, Texas A\&M University at Qatar, P.O. Box 23874 Doha, Qatar
}%

\date{\today}

\begin{abstract}
  \noindent
We investigate linear and nonlinear evolution dynamics of light beams propagating along a dislocated edge-centered square lattice.
The band structure and Brillouin zones of this novel lattice are analyzed analytically and numerically.
Asymmetric Dirac cones as well as the corresponding Bloch modes of the lattice are obtained.
By adopting the tight-binding approximation, we give an explanation of the asymmetry of Dirac cones.
By utilizing the appropriate Bloch modes, linear and nonlinear asymmetric conical diffraction is demonstrated.
We find that both the focusing and defocusing nonlinearities can enhance the asymmetry of the conical diffraction.
\end{abstract}

\maketitle

%
\section{Introduction}

Two-dimensional photonic lattices with novel spatial periodic arrangements
can serve as good platforms for manipulating the spatial behavior of light propagating through waveguide arrays formed by the lattices \cite{lederer.pr.463.1.2008,longhi.lpr.3.243.2009,kartashov.rmp.83.247.2011,garanovich.pr.518.1.2012,windpassinger.rpp.76.086401.2013}.
Nowadays, exotic photonic lattices have drawn attention around the world with the advent of photonic topological insulators \cite{rechtsman.nature.496.196.2013},
which have ushered a new discipline --- the topological photonics \cite{lu.np.8.821.2014}. It already flourishes, due to various realized or potential applications \cite{ozawa.arxiv.2018}.
The combination of topological and nonlinear properties of the material enable the formation of unidirectional topological quasi-solitons,
as predicted in waveguide arrays \cite{ablowitz.pra.90.023813.2014,leykam.prl.117.143901.2016,lumer.pra.94.021801.2016} and polaritonic systems \cite{kartashov.optica.3.1228.2016,bleu.prb.93.0854382.2016,gulevich.sr.7.1780.2017,li.prb.97.081103.2018}.
In addition to photonics, related topological topics have been frequently reported in
acoustics \cite{yang.prl.114.114301.2015,he.np.12.1124.2016}, mechanics \cite{huber.np.12.621.2016}, and physics of cold atoms \cite{beeler.nature.498.201.2013,kennedy.prl.111.225301.2013,jotzu.nature.515.237.2014,leder.nc.7.13112.2016}, to name a few.
It should also be mentioned that the lattice predominantly used is the honeycomb lattice, probably owing to the familiar hexagonal arrangement of carbon atoms, for example in graphene.
Nonetheless, the Lieb lattice \cite{leykam.pra.86.031805.2012,vicencio.prl.114.245503.2015,mukherjee.prl.114.245504.2015,xia.ol.41.1435.2016,diebel.prl.116.183902.2016}, kagome lattice \cite{vicencio.jo.16.015706.2014,zong.oe.24.8877.2016}, as well as the superhoneycomb lattice \cite{lan.prb.85.155451.2012,zhong.adp.529.1600258.2017,zhu.oe.26.24307.2018} are also becoming popular, due to their unique band topologies and interesting physical properties.

The mentioned lattices possess Dirac cones \cite{liu.pccp.15.18855.2013} in their band structures, and this feature is crucial in
the appearance of intriguing topological properties.
The topological phase transition demands the breakup of Dirac cones, by introduction of a proper breaking mechanism
(even though this is required, it is not sufficient).
The existence of Dirac cones means that the dispersion or diffraction around a Dirac cone is linear.
If the corresponding Dirac cone states are excited, due to linear dispersion relation, the incident light beam will propagate along circular rings with invariant width during propagation;
this phenomenon is known as the conical diffraction. It is a well-known mode of beam propagation,
which has been reported in different lattices, such as honeycomb, Lieb,
and super-honeycomb \cite{peleg.prl.98.103901.2007,ablowitz.pra.79.053830.2009,ramezani.pra.85.013818.2012,leykam.pra.86.031805.2012,diebel.prl.116.183902.2016,song.nc.6.6272.2015,zhong.adp.529.1600258.2017,song.cleo.2015}.
Recent research demonstrates that the conical diffraction has potential applications
in variety of areas, including optical trapping, optical communications, super-resolution imaging, lasers, and others \cite{turpin.lpr.10.750.2016}.

Until now, the reported conical diffractions were always symmetric, that is to say, the rings are circular or at most elliptic \cite{ablowitz.pra.79.053830.2009,ablowitz.pra.82.013840.2010,song.cleo.2015}, which means they are still symmetric about the $x=0$ and $y=0$ axes.
Indeed, the asymmetric conical diffraction has not been investigated yet in sufficient detail.
 Concerning conical diffraction in principle, there is no reason why Dirac cones could not be tilted \cite{trescher.prb.91.115135.2015},
which would lead to unequal speeds of the light beam along the azimuthal direction. The problem is, how to produce the tilting.
Generally, as far as we know, the Dirac cones can be classified into three types: type-I, type-II \cite{pyrialakos.prl.119.113901.2017,Charlie.nc.9.2194.2018} 
and type-III \cite{milicevic.arxiv.1807.2018}.
Type-I Dirac cone is the commonest and it leads to the symmetric conical diffraction.
The other two types are tilted, and consequently they lead to asymmetric conical diffraction.
Hence, if one introduces dislocation \cite{ferrando.apl.78.3184}, stretching, compression, or distortion,
one may obtain the tilted type-I Dirac cones, which also results in an asymmetric conical diffraction.

In this paper, we design an edge-centered square lattice with dislocation but without distortion, which possesses tilted type-I Dirac cones and of course
abundantly displays asymmetric conical diffraction.
In addition to the linear conical diffraction, the nonlinear diffraction will also be investigated in this paper.
We believe that our research will not only enrich the family of conical diffraction phenomena and develop means to observe asymmetric conical diffraction,
but also provide a new platform for investigating related topological effects.

We use a dimensionless model to carry out our investigation, but obtained phenomena can experimentally be observed in photorefractive (e.g., SBN or ${\rm LiNbO}_3 $)
or silicon-based materials.
The reason is that almost any kind of lattices and waveguide arrays can be formed in such materials, by using the
optically-induced generation method or the femto-second laser writing technique \cite{peleg.prl.98.103901.2007,rechtsman.nature.496.196.2013,plotnik.nm.13.57.2014,xia.ol.41.1435.2016,zong.oe.24.8877.2016}.

The organization of this paper is as follows.
In Sec. \ref{model}, we display the configuration of the lattice and the corresponding two-dimensional band structure.
The one-dimensional band structure of the strained lattices with different boundaries is also depicted.
Together with the one-dimensional band structure, the edge states as well as the Dirac cone states are also shown.
In Sec. \ref{linear}, we present the asymmetric conical diffraction in detail, by launching Dirac cone states into the lattice.
In Sec. \ref{nonlinear}, the nonlinear conical diffraction is discussed, by introducing both focusing and defocusing nonlinearities into the model.
We conclude the paper with Sec. \ref{conclusion}.

\section{The lattice, the band structure, and the edge and cone states} \label{model}

\begin{figure}[htpb]
  \centering \includegraphics[width=0.5\columnwidth]{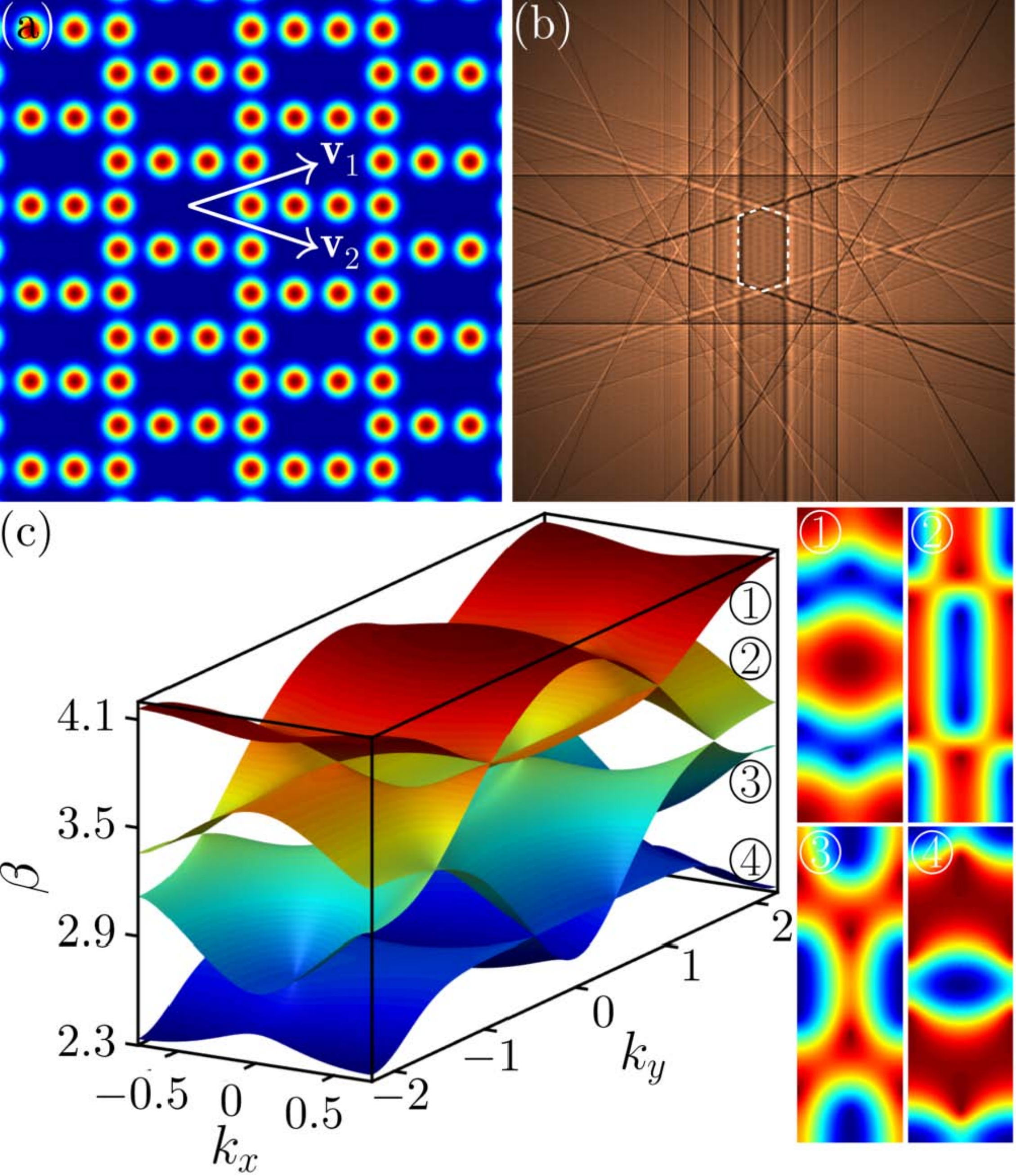}
  \caption{(a) Dislocated edge-centered square lattice.
  (b) Far-field diffraction pattern of the lattice. The dashed hexagon is the first Brillouin zone.
  (c) Band structure of the lattice.
  The right four panels are the top-views of the four bands.}
  \label{fig1}
\end{figure}

The dislocated edge-centered square lattice is shown in Fig. \ref{fig1}(a).
The lattice is constructed using Gaussian beams, $R(x,y)=\sum_{m,n}p_{\rm in}\exp[-(x-ma)^2/w^2-(y-na)^2/w^2]$,
in which $a=1.4$ is the distance between two nearest-neighbor sites,
$w=0.5$ is the width of the Gaussian beam,
and $p_{\rm in}=10$ is the potential depth.
The corresponding unit vectors of the lattice are ${\bf v}_1=[3a,a]$ and ${\bf v}_2=[3a,-a]$,
which means that there are 4 sites in one unit cell.
The far-field diffraction pattern \cite{bartal.prl.94.163902.2005} of this lattice is displayed in Fig. \ref{fig1}(b),
where the white dashed elongated hexagon denotes the first Brillouin zone.
To see the band structure of the lattice, we have to provide for
the propagation of light beams in this lattice; this can be conveniently described by the dimensionless Schr\"odinger-like paraxial wave equation
\begin{equation}\label{eq1}
i\frac{\partial \psi(x,y,z)}{\partial z} + \frac{1}{2} \left(\frac{\partial^2}{\partial x^2}+\frac{\partial^2}{\partial y^2}\right) \psi(x,y,z) + R(x,y) \psi(x,y,z) + g|\psi(x,y,z)|^2\psi(x,y,z)=0,
\end{equation}
in which $\psi(x,y,z)$ stands for the slowly-varying envelope of the light beam and $g=\pm 1$ picks the focusing/defocusing Kerr nonlinearity.
Thus, the beams propagate along the $z$ direction, while the lattice is formed in the transverse ($x,y$) plane.

We first consider the linear case with $g=0$.
To ease mathematical analysis, we first rotate the lattice into the frame $(x',y')$
by an angle $\theta=\arctan(1/3)$ anticlockwise, {\it i.e.}, $(x,y)\xrightarrow{\quad\theta\quad}(x',y')$,
and then stretch the lattice, to make the two unit vectors perpendicular to each other via the relations $x'=X+Y\cos(2\theta)$ and $y'=Y\sin(2\theta)$,
where $\sin(2\theta)=3/5$ and $\cos(2\theta)=4/5$.
Hence, in the new frame $(X,Y)$, Eq. (\ref{eq1}) can be rewritten as
\begin{equation}\label{eq2}
i\frac{\partial \psi(X,Y,Z)}{\partial Z} + \frac{5}{18} \left(5\frac{\partial^2}{\partial X^2}+5\frac{\partial^2}{\partial Y^2}-8\frac{\partial^2}{\partial X \partial Y}\right) \psi(X,Y,Z) + R(X,Y) \psi(X,Y,Z)=0.
\end{equation}
Note that the longitudinal coordinate is not affected by the transformation, so $Z=z$.
Based on the plane-wave expansion method \cite{zhang.lpr.10.526.2016},
the band structure can be easily obtained, as displayed in Fig. \ref{fig1}(c) and the four panels nearby.
In the band structure, the Bloch momenta $k_x$ and $k_y$ correspond to the frame $(x,y)$.
One finds that indeed there are 6 Dirac cones between the two neighboring bands,
but they are tilted and elliptic, and are not located at the 6 corners of the first Brillouin zone.
This is significantly different from the honeycomb lattice \cite{peleg.prl.98.103901.2007,song.nc.6.6272.2015}.
As a result, when Dirac cone modes are excited,
one will observe an asymmetric conical diffraction.

Owing to the mathematical complexity of transformations,
we aim at utilizing an easy way to
excite the Dirac cone modes.
The simplest is to use Gaussian beams superposed with the momenta belonging to the tilted and elliptic Dirac cones \cite{ablowitz.pra.79.053830.2009}.
However, the Gaussian beams will inevitably excite the bulk modes as well and lead to radiation mixed with the conical diffraction,
because the locations of the Dirac cones in the three band gaps are different. So, one should pay close attention to how the obtained modes are interpreted.

\begin{figure}[htpb]
\centering \includegraphics[width=1\columnwidth]{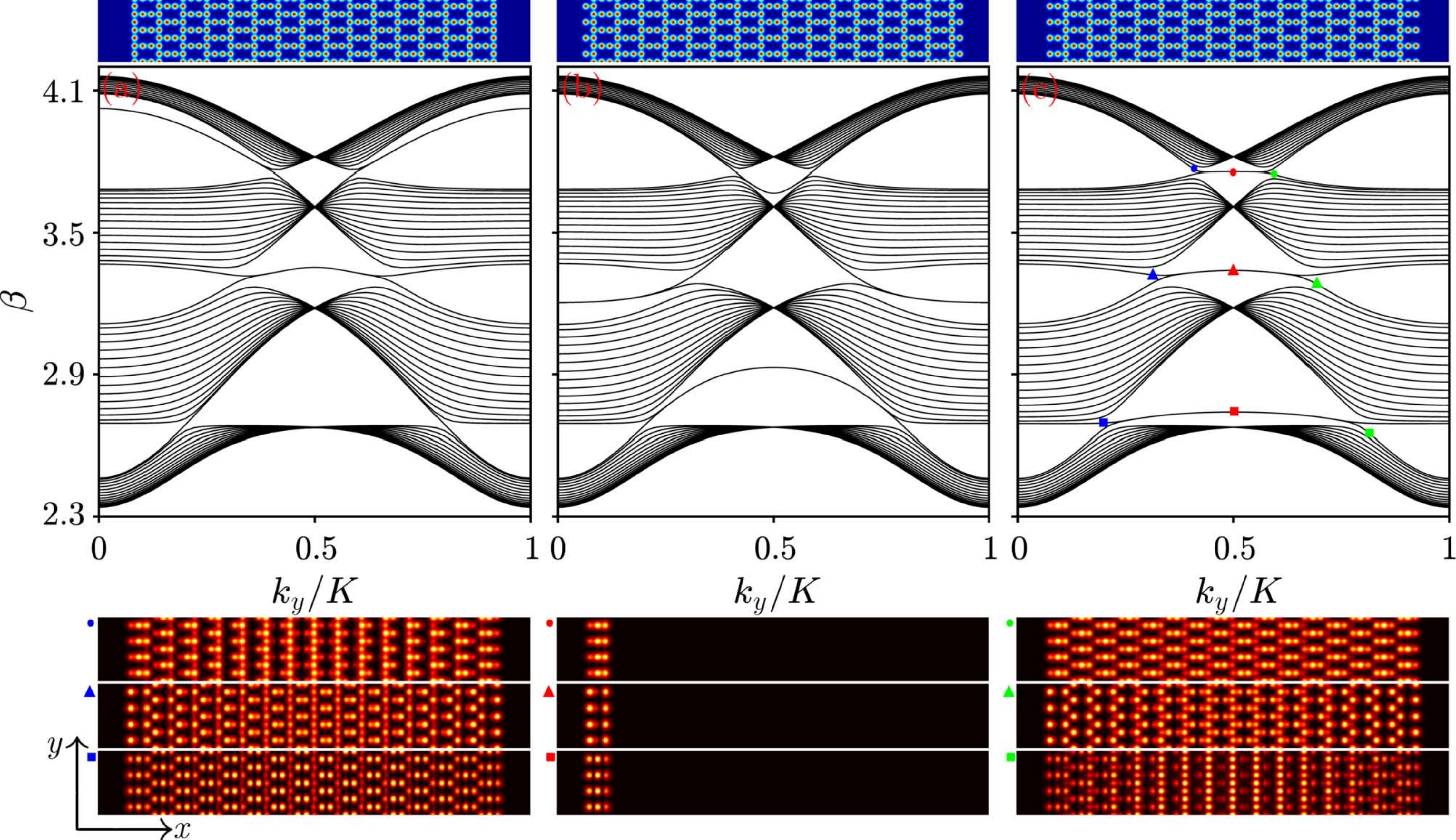}
\caption{(a) Band structure of the lattice strained with flat boundaries along $x$ direction and periodic along $y$ direction.
(b) and (c) Same as (a) but with short-bearded boundaries and long-bearded boundaries along $x$ direction. The corresponding lattices are shown above the band structure.
The states below the band structures correspond to the edge and cone states marked with different labels in (c).
The lattices and states are shown in the window $-38\le x \le 38$ and $-4d \le y \le 4d$.
}
\label{fig2}
\end{figure}

If the lattice is strained along $x$ direction but still periodic along $y$ direction,
one may consider the three interesting cases: lattices with
\begin{enumerate}
  \item[(i)] flat boundaries
  \item[(ii)] short-bearded boundaries, and
  \item[(iii)] long-bearded boundaries.
\end{enumerate}
The corresponding band structures are shown in Figs. \ref{fig2}(a)-\ref{fig2}(c),
with the corresponding lattice configuration displayed above each band structure.
Here, $K=2\pi/D_y$ with $D_y=2a$ being the period along $y$ direction.
One finds that for each case, there are edge states in the band gap,
but the edge states are located at different places in the first Brillouin zone.
Of special interest is the case in Fig. \ref{fig2}(c),
in which the edge states are always located at the boundary of the first Brillouin zone.
In Fig. \ref{fig2}(c), we select some characteristic states that are marked with dots, triangles, and squares of different colors.
Corresponding to these points, the states are shown in Fig. \ref{fig2}, below the band structures.
Clearly, the states with the red marks are the edge states, so the energy there is mostly located along the edge of the lattice.
The states labeled with blue and green marks are located around the Dirac cones,
so they are close to the Dirac cone states.
One can use these states to track the formation of conical diffraction in the lattice. Different possibilities are analyzed in the following section.

\section{Linear conical diffraction}\label{linear}

\begin{figure}[htpb]
\centering \includegraphics[width=1\columnwidth]{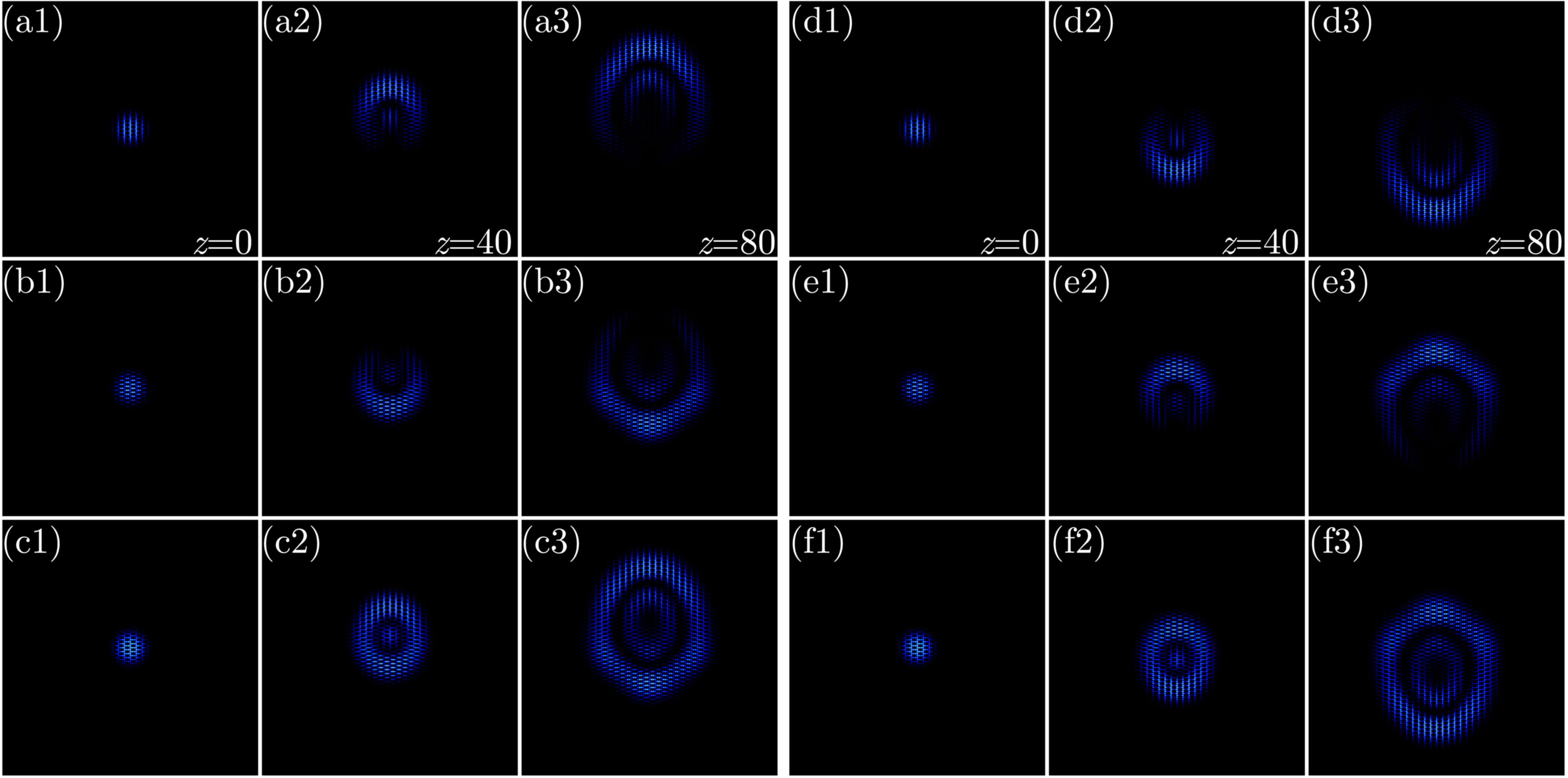}
\caption{(a) Conical diffraction of the state marked with a blue dot in Fig. \ref{fig2}(c) superposed by a Gaussian, at selected propagation distances.
(b) Same as (a), but the state is that below the blue dot in Fig. \ref{fig2}(c).
(c) Same as (a), but with both the states used in (a) and (b) being excited.
(d) Conical diffraction of the state above the green dot in Fig. \ref{fig2}(c).
(e) Same as (d), but the state is marked by the green dot in in Fig. \ref{fig2}(c).
(f) Same as (d), but with both states used in (d) and (e) being excited.
The panels are shown in the window $-90\le x \le 90$ and $-90 \le y \le 90$.
}
\label{fig3}
\end{figure}

At the beginning, we propagate the state marked with a blue dot in Fig. \ref{fig2}(c),
and Figs. \ref{fig3}(a1)-\ref{fig3}(a3) present the intensity of the beam at selected distances.
One finds that the conical diffraction indeed appears, but there is a dark notch at the bottom.
The conical diffraction is asymmetric: the spreading speed along the positive $y$ direction is much larger than that along the negative $y$ direction,
and the speeds along both the positive and negative $x$ directions are the same.
On the other hand, if the propagation of the state on the line just below the blue dot is considered, the
results look mirror-image, as displayed in Figs. \ref{fig3}(b1)-\ref{fig3}(b3).
The dark notch now appears at the top of the asymmetric conical diffraction
and the intensity is inverted.
The appearance of the notches in Figs. \ref{fig3}(a1)-\ref{fig3}(a3) and \ref{fig3}(b1)-\ref{fig3}(b3) is due to
the fact that only one of the two Dirac cone states are excited in each mode,
which is different from the notch appearance in a honeycomb lattice reported previously \cite{ablowitz.pra.79.053830.2009,song.nc.6.6272.2015}.
If Gaussian beams are used to excite the Dirac states,
a pair of states about the blue dot could be excited simultaneously and there will be no notch in the conical diffraction.
This is visible in Figs. \ref{fig3}(c1)-\ref{fig3}(c3), where the two Dirac cone states are used together as the input.

Similar to the honeycomb lattice, which possesses different $\bf K$ and ${\bf K}'$ Dirac cones,
the dislocated edge-centered square lattice also possesses different cones.
If the Dirac cone marked with blue dot is $\bf K$, then the Dirac cone marked with green dot is ${\bf K}'$.
As a comparison, we show the conical diffractions corresponding to the ${\bf K}'$ Dirac cone in Figs. \ref{fig3}(d1)-\ref{fig3}(d3), \ref{fig3}(e1)-\ref{fig3}(e3),
and \ref{fig3}(f1)-\ref{fig3}(f3).
One finds that the formation of conical diffraction related to the ${\bf K}'$ Dirac cone can be viewed as a mirroring process
of that related to the ${\bf K}$ Dirac cone, about $y=0$.

In order to unveil the tilted Dirac cones, we adopt the tight-binding approximation method by considering the nearest-neighbor interaction.
The corresponding Hamiltonian can be written as
\begin{equation}\label{eq3}
{\cal H}=
\begin{bmatrix}
0 & \exp(i{\bf k}\cdot {\bf e}_1) & 0 & 2\cos(i{\bf k}\cdot {\bf e}_2) \\
\exp(-i{\bf k}\cdot {\bf e}_1) & 0 & \exp(i{\bf k}\cdot {\bf e}_1) & 0 \\
0 & \exp(-i{\bf k}\cdot {\bf e}_1) & 0 & \exp(i{\bf k}\cdot {\bf e}_1) \\
2\cos(i{\bf k}\cdot {\bf e}_2) & 0 & \exp(-i{\bf k}\cdot {\bf e}_1) & 0
\end{bmatrix},
\end{equation}
where ${\bf e}_1=[a,0]$, ${\bf e}_2=[0,a]$, and ${\bf k}=[k_x,k_y]$.
According to Eq. (\ref{eq3}), the location of the ${\bf K}'$ Dirac point of the first band is at $[k_x=0,k_y=2\pi/3a]$.
One can expand the Hamiltonian about the Dirac point, calculate the eigenvalues and after some approximations, obtain
the perturbed band around the Dirac point, as described by
\begin{equation}\label{eq4}
\beta=\sqrt{2}+\frac{\sqrt{6}}{4}ap_y+\frac{\sqrt{2}}{4}a\sqrt{9p_x^2+6p_y^2},
\end{equation}
where $p_{x,y}$ are small momenta.
Clearly, the third term in Eq. (\ref{eq4}) is an elliptic cone,
and the second term will make the elliptic cone tilted.
We should note that Eq. (\ref{eq4}) is obtained based on the tight-binding approximation,
and this result does not fully agree with that based on the continuous model,
but it still reflects on the asymmetric conical diffraction qualitatively.
It should also be noted that due to Eq. (\ref{eq4}), the Dirac cone is always tilted in the same way; 
however, if one changes the depth of the lattice site or introduces some distortion to the lattice,
the Dirac cone may be tilted in a different way ---  the ratio between the second and third terms in Eq. (\ref{eq4}) may change.

\begin{figure}[htpb]
\centering \includegraphics[width=1\columnwidth]{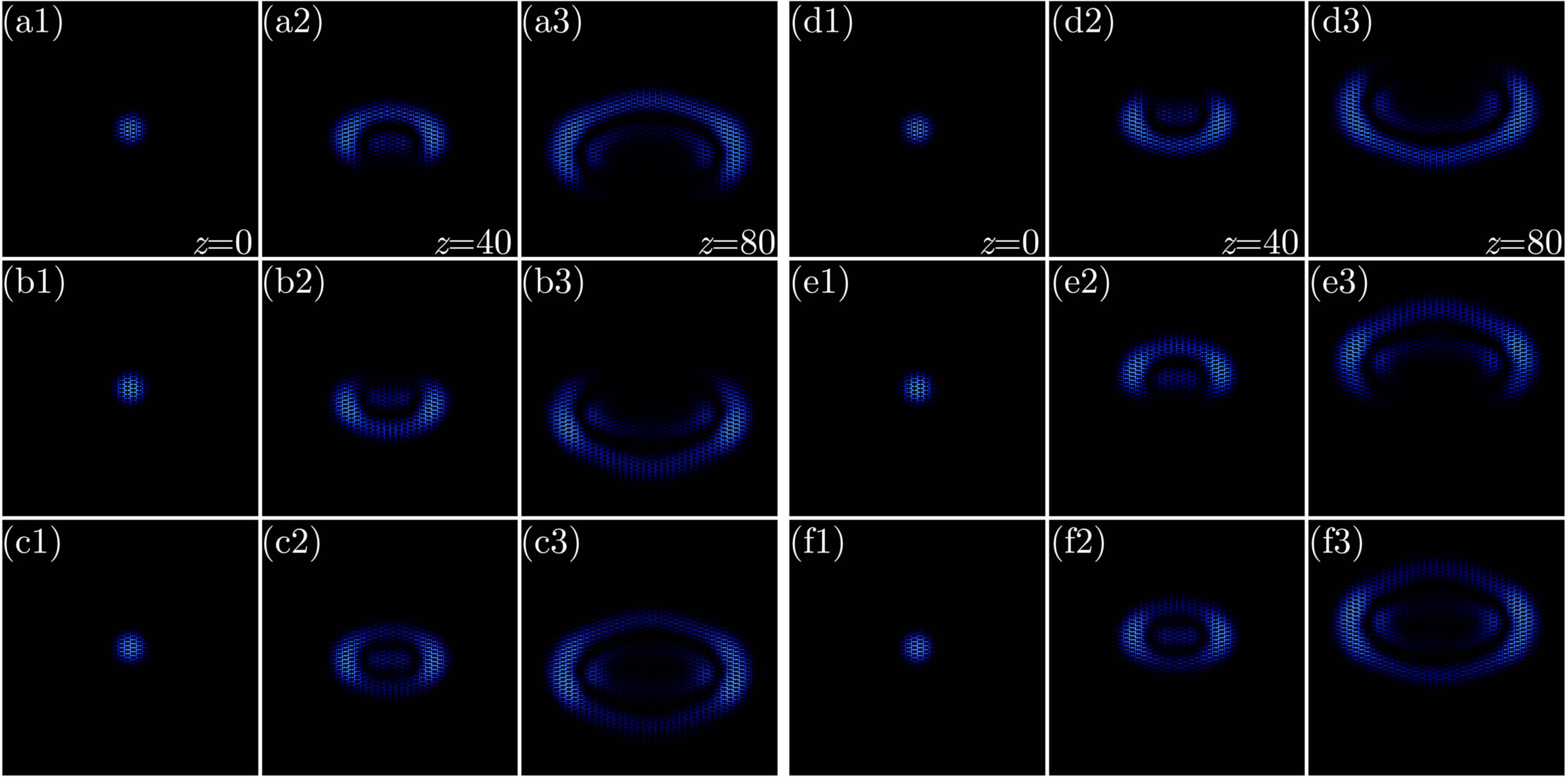}
\caption{Figure setup is as Fig. \ref{fig3}, but for the states marked with triangles in Fig. \ref{fig2}.}
\label{fig4}
\end{figure}

\begin{figure}[htpb]
\centering \includegraphics[width=1\textwidth]{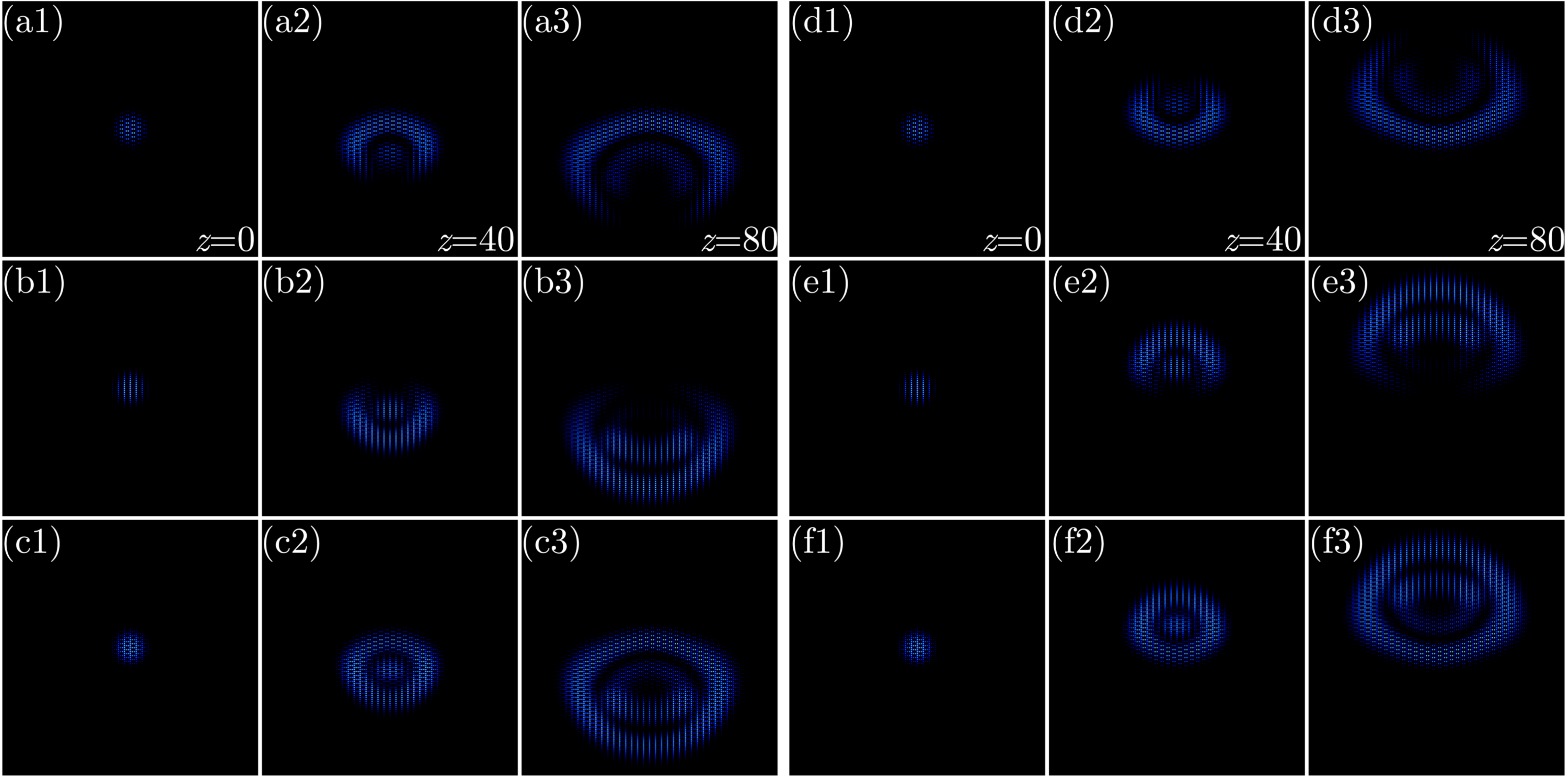}
\caption{Figure setup is as Fig. \ref{fig3}, but for the states marked with squares in Fig. \ref{fig2}.}
\label{fig5}
\end{figure}

As concerns the Dirac cones in the second and third band gaps, we investigate the corresponding conical diffraction by taking a similar procedure,
and the results are shown in Figs. \ref{fig4} and \ref{fig5}.
And similarly, asymmetric conical diffraction is also obtained.
Different from Fig. \ref{fig3}, where the spreading speed is bigger along the $y$ direction than that along the $x$ direction,
the spreading speed in Figs. \ref{fig4} and \ref{fig5} is bigger along the $x$ direction.
It is also worth mentioning that the conical diffraction almost does not spread along the positive $y$ axis
in Figs. \ref{fig4}(a)-\ref{fig4}(c) and \ref{fig5}(a)-\ref{fig5}(c), or does not spread along the negative $y$ axis
in Figs. \ref{fig4}(d)-\ref{fig4}(f) and \ref{fig5}(d)-\ref{fig5}(f).
The reason is that
one side of the Dirac cones in the second and third band gaps is almost flat, which corresponds
to a nearly-zero moving velocity of the state along this direction.
Frankly speaking, the rings of the conical diffraction will also change from circular to elliptic if the
deformation is introduced into the honeycomb lattice \cite{ablowitz.pra.82.013840.2010},
but this kind of conical diffraction is still symmetric about $x=0$ as well as about $y=0$.
Therefore, what we discovered here is completely different, with tilted and elliptic Dirac cones appearing without any deformation or stretching.

\begin{figure}[htpb]
\centering \includegraphics[width=0.625\columnwidth]{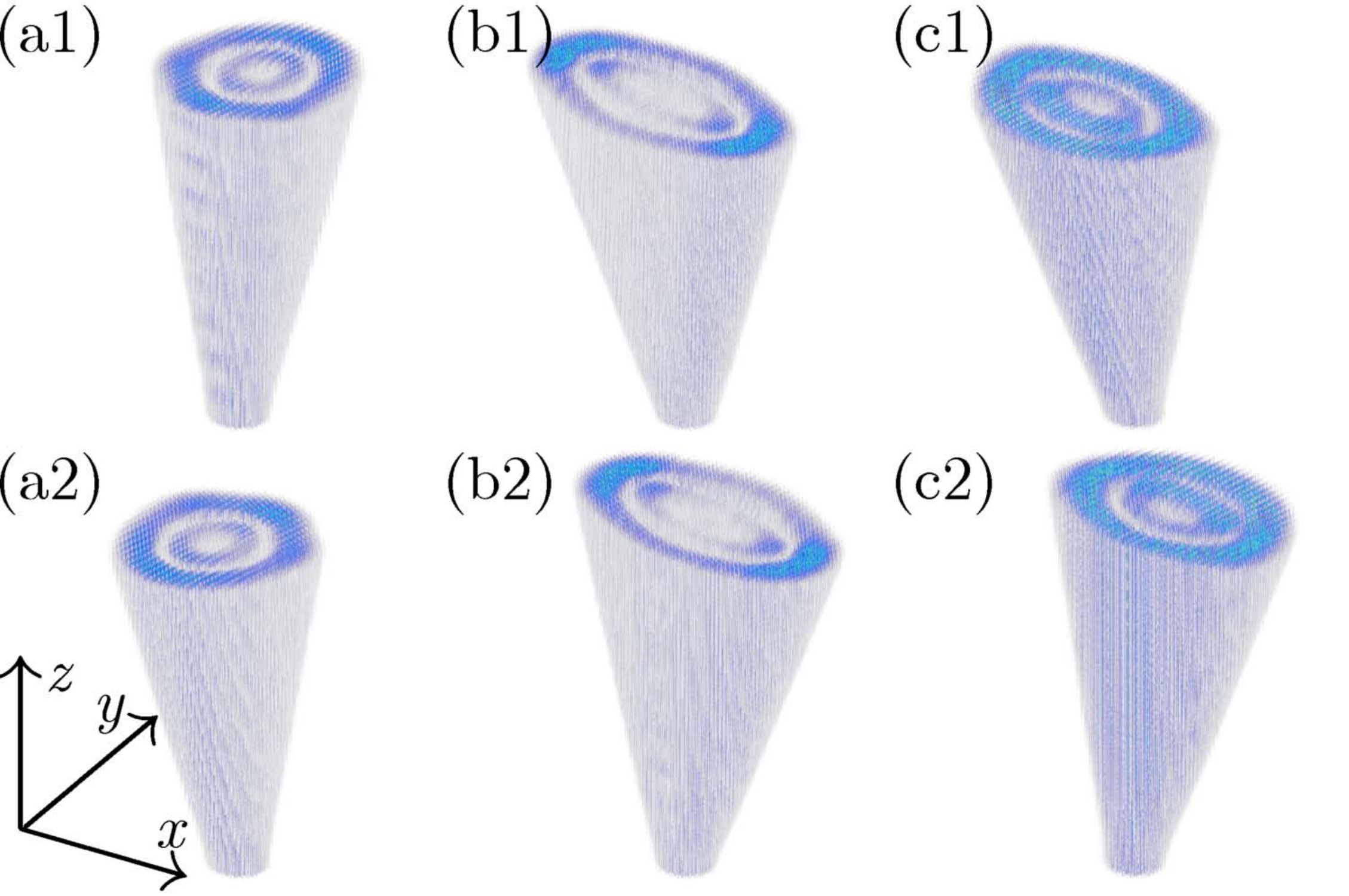}
\caption{Three-dimensional formation process of the asymmetric conical diffraction.
(a1) and (a2) correspond to Figs. \ref{fig3}(c3) and \ref{fig3}(f3);
(b1) and (b2) correspond to Figs. \ref{fig4}(c3) and \ref{fig4}(f3);
(c1) and (c2) correspond to Figs. \ref{fig5}(c3) and \ref{fig5}(f3).
The visual angles for these panels are same.
}
\label{fig6}
\end{figure}

To see the asymmetric conical diffractions more obviously,
we depict the three-dimensional formation process of the propagation of the states shown in Fig. \ref{fig6}.
From the process, one finds that the size of the asymmetric diffraction rings indeed
increases linearly with the propagation distance.

\section{Nonlinear conical diffraction} \label{nonlinear}

The nonlinear conical diffraction has been reported in the previous literature \cite{ablowitz.pra.79.053830.2009,ablowitz.ol.36.3762.2011,leykam.pra.86.031805.2012}.
However, this diffraction was still symmetric; the truly nonlinear asymmetric conical diffraction has not been reported yet. It is reported here, for the first time.

\begin{figure}[htpb]
\centering \includegraphics[width=1\textwidth]{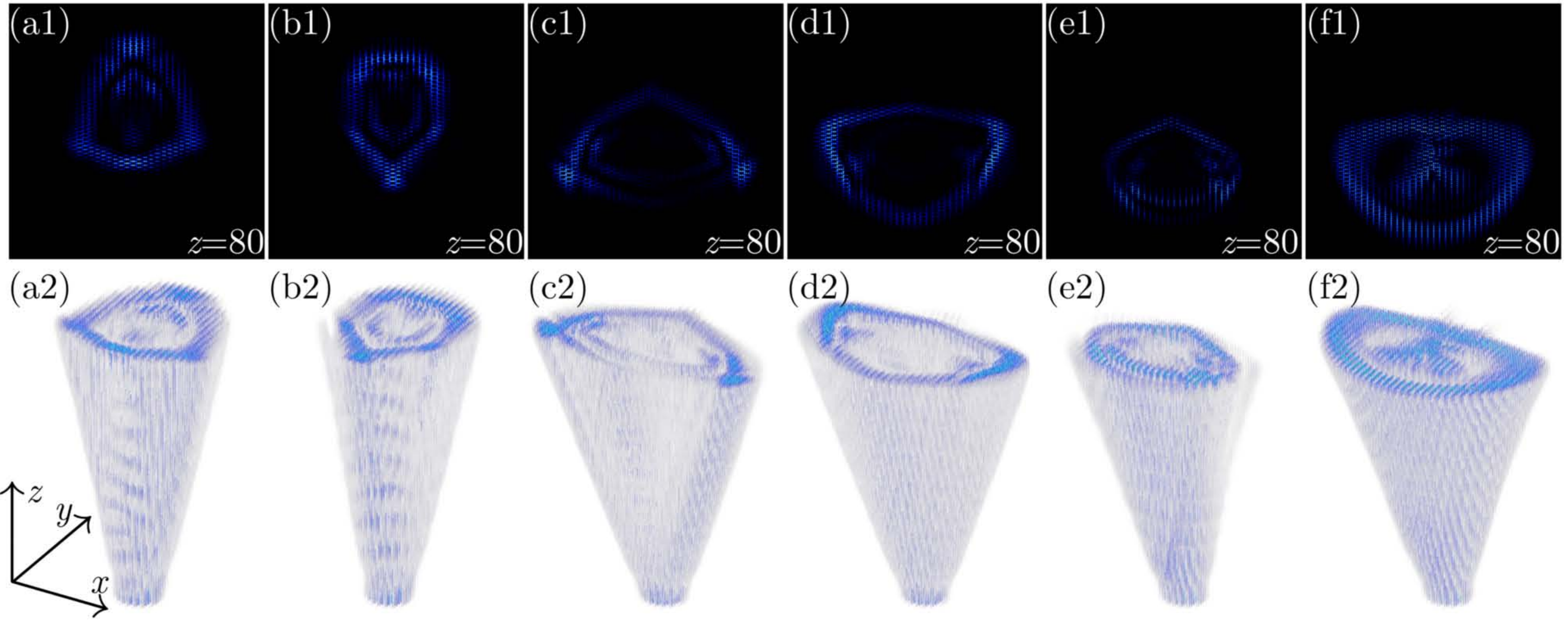}
\caption{Nonlinear conical diffraction. (a1), (c1), (e1) The focusing case $g=1$, corresponding to Figs. \ref{fig3}(c3), \ref{fig4}(c3), and \ref{fig5}(c3).
(b1), (d1), (f1) The defocusing case $g=-1$, corresponding to Figs. \ref{fig3}(f3), \ref{fig4}(f3), and \ref{fig5}(f3).
The input beam power in (a1) and (b1) is 8.9, in (c1) and (d1) the beam power is 10.7, in (e1) the beam power is 10.0, and in (f1) the beam power is 13.3.
(a2)-(f2) The three-dimensional picture of the nonlinear asymmetric conical diffraction, corresponding to (a1)-(f1).}
\label{fig7}
\end{figure}

The propagation is still based on Eq. (\ref{eq1}), but with nonzero $g$.
The incident beam is a combination of Bloch modes superposed by Gaussian beams, which is the same as in Figs. \ref{fig3}(c1)-\ref{fig5}(c1) for different band gaps.
We first consider the case corresponding to Fig. \ref{fig3}(c1) under the focusing nonlinearity ($g=1$), and the result is displayed in Fig. \ref{fig7}(a1).
In comparison with the linear case in Fig. \ref{fig3}(c3), the profile of the asymmetric conical diffraction changes greatly ---
the spreading speed along $x$ direction is enhanced, looking more triangle-like.
On the other hand, for the defocusing nonlinearity ($g=-1$), as shown in Fig. \ref{fig7}(b1),
the spreading behavior along both $x$ and $y$ directions is completely different.
In the defocusing case, the speed along the negative $y$ direction becomes bigger,
while that along the $x$ and positive $y$ directions is smaller,
which makes the beam exhibit an inverted-bottle-like profile.

The conical diffraction in the second band gap as shown in Fig. \ref{fig4}(c1) or \ref{fig6}(b1), exhibits elliptical rings.
But under the nonlinear condition, the elliptic rings in Fig. \ref{fig4}(c3) change into semicircular rings, as shown in Figs. \ref{fig7}(c1) and \ref{fig7}(d1).
The difference between the focusing and defocusing cases is that the former condition prohibits the spreading along the negative $y$ direction [Fig. \ref{fig7}(c1)],
and the latter condition has an opposite effect [Fig. \ref{fig7}(d1)].
So, the bottom and top edges of the nonlinear conical diffraction in Figs. \ref{fig7}(c1) and \ref{fig7}(d1) are almost straight.
But, the full profiles in Figs. \ref{fig7}(c1) and \ref{fig7}(d1) are not mirror images of each other.

In comparison with the linear case in Fig. \ref{fig4}(c3), the nonlinear conical diffraction in Figs. \ref{fig7}(c1) and \ref{fig7}(d1) does not spread along the positive $y$ direction too.
Such a property is also conserved for the nonlinear conical diffraction in the third band gap, as presented in Figs. \ref{fig7}(e1) and \ref{fig7}(f1).
For the focusing nonlinearity [Fig. \ref{fig7}(e1)], the conical diffraction is prohibited greatly, since the asymmetric diffraction rings are
much smaller than those of the defocusing nonlinearity [Fig. \ref{fig7}(f1)].
The profile in Fig. \ref{fig7}(f1) is also semicircular, which is similar to that in Fig. \ref{fig7}(d1), but
the defocusing nonlinearity strengthens the diffraction a bit,
because the linear diffraction rings in Fig. \ref{fig5}(c3) are smaller than those in Fig. \ref{fig4}(c3). However, the nonlinear diffraction rings are almost the same in Figs. \ref{fig7}(d1) and \ref{fig7}(f1).

Based on Fig. \ref{fig7}, one finds that the focusing and defocusing nonlinearities play different but not simply opposite roles in the formation of conical diffraction.
In addition, the nonlinearity can make the asymmetric conical diffraction more asymmetric.
To clarify the whole propagation processes more clearly, similar to Fig. \ref{fig6},
we show the three-dimensional picture of nonlinear conical diffraction in Fig. \ref{fig7}(a2)-\ref{fig7}(f2).
One finds that the size of the nonlinear diffraction rings also increases with propagation distance linearly ---
a general requirement for the conical diffraction.

\section{Conclusion} \label{conclusion}

In summary, we have investigated both the linear and nonlinear asymmetric conical diffraction in the dislocated edge-centered square lattice,
which possesses four bands and six Dirac cones between each nearest two bands.
The tight-binding approximation demonstrates that the Dirac cones are asymmetric.
The Bloch modes of the asymmetric Dirac cones are obtained using the one-dimensional band of the strained lattice.
By adopting the Bloch modes superposed with Gaussian beams, asymmetric conical diffraction is found.

We also demonstrated the nonlinear asymmetric conical diffraction.
Both focusing and defocusing nonlinearies can increase the asymmetry of the diffraction,
but their influence is not simply in the opposite directions.
Our research not only reports how to observe asymmetric conical diffractions,
but also provides a new platform for studying topological effects in two-dimensional lattices of waveguides.

\section*{Funding}
Natural Science Foundation of Shaanxi Province (2017JZ019);
Natural Science Foundation of Guangdong Province (2018A0303130057);
Qatar National Research Fund (NPRP 8-028-1-001).

\section*{Acknowledgments}
MRB acknowledges support by the Al-Sraiya Holding group.

\bibliography{my_refs_library}

\end{document}